\documentclass[twocolumn,showpacs,prb,preprintnumbers,amsmath,amssymb]{revtex4-1}

\usepackage{graphicx} 
\usepackage{dcolumn}
\usepackage{bm}
\usepackage{amsmath}

\begin{document}

\title{Asymmetric optical nuclear spin pumping in a single uncharged quantum dot}  

\author{F. Klotz, V. Jovanov, J. Kierig, E. C. Clark, M. Bichler, G. Abstreiter, M. S. Brandt, and J. J. Finley}
\affiliation{Walter Schottky Institut, Technische Universit\"{a}t M\"{u}nchen, Am Coulombwall 3, 85748 Garching, Germany}

\author{H. Schwager and G. Giedke}
\affiliation{Max-Planck-Institut f\"ur Quantenoptik, Hans-Kopfermann-Stra\ss e 1, 85748 Garching, Germany}

\begin{abstract}
\noindent Highly asymmetric dynamic nuclear spin pumping is observed in a single self-assembled InGaAs quantum dot subject to resonant optical excitation of the neutral exciton transition. A large maximum polarization of 54\% is observed and the effect is found to be much stronger upon pumping of the higher energy Zeeman level. Time-resolved measurements allow us to directly monitor the buildup of the nulcear spin polarization in real time and to quantitatively study the dynamics of the process. A strong dependence of the observed dynamic nuclear polarization on the applied magnetic field is found, with resonances in the pumping efficiency observed for particular magnetic fields. We develop a model that accounts for the observed behavior, where the pumping of the nuclear spin system is due to hyperfine-mediated spin flip transitions between the states of the neutral exciton manifold. 

\end{abstract}

\maketitle

Nuclear spin effects in semiconductor quantum dot (QD) nanostructures have attracted much attention over recent years. \cite{Coi09, TML03, Kha02, Oul07, Xu09, MBI07, Ebl06, Bra05, GEK+01, Ta07, YAS+05, Mo09, MKI08, Mak08} The hyperfine (hf) interaction of the $10^4$-$10^5$ nuclear spins within the dot and the spin of an individual electron that is electrically or optically generated is key to address and control the nuclear spin system. This may provide opto-electronic access to the mesoscopic nuclear spin system with strong potential for future applications in quantum information technologies.\cite{TML03} The hf interaction limits the electron spin coherence in QDs (Refs. 3 and 15) making reliable strategies to control the nuclear field highly desirable.\cite{Oul07, Xu09} From both perspectives, a highly polarized ensemble of nuclear spins would be advantageous. To date, the vast majority of experiments on dynamical nuclear polarization (DNP) have been carried out on charged QDs containing a resident electron.\cite{MBI07, Ebl06, Bra05} However, this system is subject to fast depolarization effects that are typically mediated by the residual electron in the dot.\cite{MBI07} Comparatively few studies of DNP in neutral QDs  have been reported,\cite{GEK+01, Ta07, YAS+05, Mo09, MKI08, Mak08, Kor99} for which a stable polarization of the nuclear spin system over timescales exceeding 1 h has been demonstrated.\cite{MKI08} 

In this Rapid Communication we demonstrate pumping of the nuclear spin system in an InGaAs QD via resonant optical excitation of the neutral exciton X$^0$. Most surprisingly, a strong asymmetry is found in the DNP efficiency for excitation of the two transitions of the orbital ground state of the dot. DNP predominantly occurs for pumping of the higher energy Zeeman level. A theoretical model is presented that accounts for the experimental results. Our results show that the asymmetric and efficient resonant DNP arises from hf-mediated spin flip transitions between neutral exciton states. 

\begin{figure}[b]
	\centering
		\includegraphics[width=0.42\textwidth]{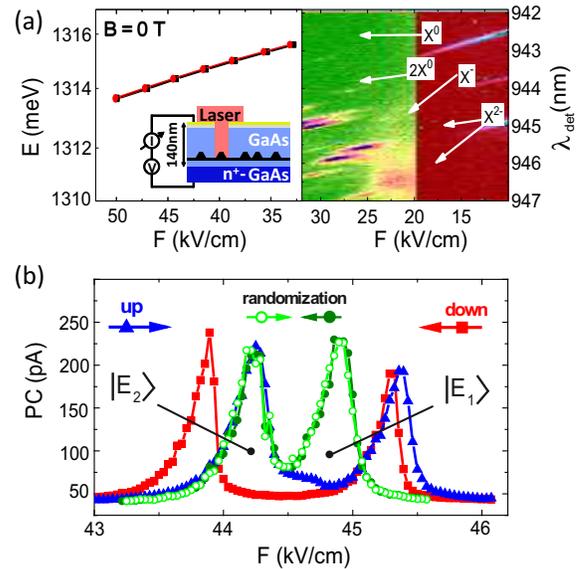}
	\label{fig:plot001}
	\caption{(colour online) (a) Combined PC and PL measurements at $B_{\mathrm{ext}}=$ 0T. Inset: Schematic of the structure investigated consisting of a single layer of self assembled QDs is embedded in the intrinsic region of Schottky photodiode.(b) Both X$^0$ $s$-shell states measured in PC electric field sweeps at $B_{\mathrm{ext}}=$ 5T. The individual curves were measured as described in the text.}
\end{figure}

As depicted schematically in the inset of Fig.~1a, the samples investigated were GaAs \textit{n}-\textit{i}-Schottky photodiode structures grown by molecular beam epitaxy. A single layer of nominally In$_{0.5}$Ga$_{0.5}$As self-assembled QDs was grown in the i-region using a partially covered island growth mechanism. The diode is formed by a heavily \textit{n}$^+$-doped back contact and a 3nm thick semitransparent Ti top contact which allows the application of dc electric fields along the growth direction of the QDs. The Ti top contact is covered with an opaque Au layer in which 1$\mu$m wide circular apertures are opened to facilitate optical access to single QDs. Photocurrent (PC) measurements were carried out on this structure at 10K for different magnetic fields $B_{\mathrm{ext}}$ using linearly polarized light from a tunable external cavity Littman-Metcalf diode laser. We employ the quantum confined Stark effect (QCSE) to tune the transitions of the QD into resonance with the laser by sweeping the applied electric field while keeping the laser energy fixed. Figure 1a shows the dc Stark shift of the examined X$^0$ state measured at $B_{\mathrm{ext}}=0$T in both PC at high ($>30$kV/cm) and photoluminescence (PL) at lower electric fields ($<30$kV/cm) which can be well described using a second order polynomial fit allowing a direct conversion of applied electric field into transition energy.\cite{Fry00}
Figure 1b shows an example of an electric field sweep PC measurement performed at $B_{\mathrm{ext}}=5$T. The measurement clearly reveals the two optically active (bright) $s$-shell states of X$^0$, denoted $\left| E_1 \right\rangle$ and $\left| E_2 \right\rangle$, as they are tuned into and out of resonance with the laser by the QCSE. The levels $\left| E_1 \right\rangle$ and $\left| E_2 \right\rangle$ are separated by an energy gap $\Delta E = \sqrt{E_{\mathrm{Z}}^2 + \delta_1^2}$ in an externally applied magnetic field, where $E_{\mathrm{Z}}$ is the Zeeman energy and $\delta_1$ the fine structure splitting due to anisotropic exchange coupling.\cite{Bay02} For $E_{\mathrm{Z}}  \gg \delta_1$, the states $\left| E_{1,2} \right\rangle$ correspond to the bright excitons with angular momentum projection $J_z=+$1 ($ \left| \downarrow \Uparrow \right\rangle$) and $J_z=-$1 ($\left| \uparrow \Downarrow \right\rangle$), respectively, where $\uparrow$, $\downarrow$ ($\Uparrow,\Downarrow$) denote the electron (hole) spin orientation.    
A clear difference is observed between the two measurements performed with opposite sweep directions of the electric field; from low to high values ('sweep up' - blue trace, closed triangles in Fig.~1b) and high to low values ('sweep down' - down"—red trace, closed squares in Fig.~1b). These observations are shown to arise from DNP and the resulting effective Overhauser magnetic field $B_{\mathrm{N}}$. Partial polarization of the nuclear spin bath in the QD arises from hf coupling to the spins of the electrons pumped through the dot during the measurement and introduces an Overhauser energy shift $\delta_{\mathrm{N}} = g_{\mathrm{e}} \mu_{\mathrm{B}} B_{\mathrm{N}}$, where $\mu_{\mathrm{B}}$ is the Bohr magneton and $g_{\mathrm{e}}$ the electron 
$g$-factor. 
To study DNP it is important to obtain a reference measurement for an unpolarized state of the QD nuclear spin system. We obtain such data by first randomizing the nuclear spins prior to every measurement point recorded during an electric field sweep and also ensuring that our measurement does not induce significant DNP. To achieve this, the sample was tuned close to flatband and excited non-resonantly in the wetting layer for 10s with linearly polarized light. This procedure pumps randomly oriented electron spins through the QD, whereupon hf interactions efficiently depolarize the nuclear spin system. This expectation is confirmed by the observations presented in Fig.~1b; when applying this randomization procedure the sweep direction is found to have no influence on the measured resonance curves, the light (open circles) and dark (closed circles) green traces in Fig.~1b corresponding to sweep up and sweep down directions, respectively. We note that the observed insensitivity to sweep direction when employing the randomization procedure is in strong contrast to the results obtained without randomization during the sweep (red trace, closed squares and blue trace, closed triangles in Fig.~1b).
For all measurements without randomization, the electric field sweep was performed at a speed that was slow compared to the time required to reach the steady state polarization of the nuclear spin system. Therefore, each of the measurement points presented in Fig.~1b represents the steady state situation of the nuclear spin system. However, before every sweep we applied the randomization process once to ensure a well defined initial state without residual DNP. 

\begin{figure}[t]
	\centering
		\includegraphics[width=0.44\textwidth]{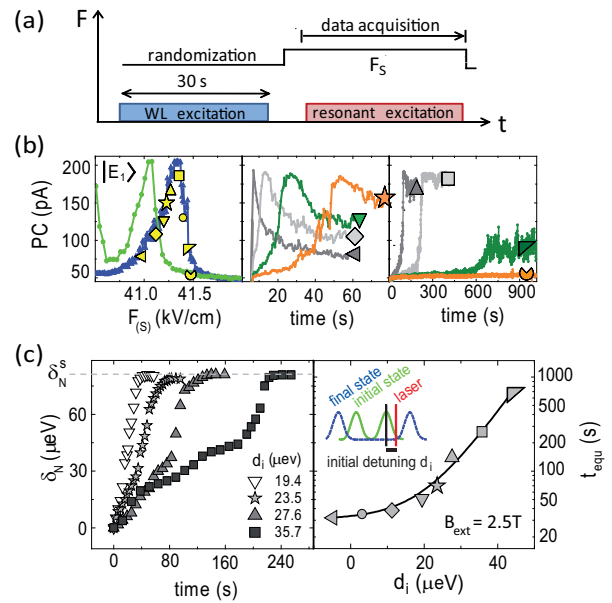}
	\label{fig:plot004}
	\caption{(colour online) (a) Timing of electric field sequence and laser excitation of the sample used for time resolved measurements of $\delta_{\mathrm{N}}$. (b) Center and right panel: Time dependent PC signal of $\left| E_1 \right\rangle$ for different initial detunings of the spin state from the laser $d_{\mathrm{i}}$ (defined by $F_{\mathrm{S}}$). Left panel: Steady state PC values of the time dependent measurements as a function of $F_{\mathrm{S}}$ (yellow/light gray symbols) in comparison with corresponding electric field sweep measurements (green trace, closed circles and blue trace, closed triangles; color and symbol coding identical to Fig. 1b). (c) Left panel: $\delta_{\mathrm{N}}^{\mathrm{s}}$ as a function of time for different $d_{\mathrm{i}}$. Right panel: Time $t_{\mathrm{equ}}$ until $\delta_{\mathrm{N}}^{\mathrm{s}}$ is reached as a function of $d_{\mathrm{i}}$. The  black solid line is a exponential fit where an offset of 30s is included. Identical symbols ($\triangle, \nabla$,...) in different panels refer to the same $d_{\mathrm{i}}$.}
\end{figure}

Figure 1b reveals an asymmetric behavior of DNP upon pumping of the two bright exciton states of X$^0$. For the up-sweep (blue trace, closed triangles in Fig.~1b), first $\left| E_2 \right\rangle$ comes into resonance with the laser as the energies of the states are shifted via the QCSE. No significant DNP effects are observed upon exciting $\left| E_2 \right\rangle$ since the measured PC signal coincides exactly with the reference curve recorded without DNP effects. However, as $\left| E_1 \right\rangle$ is tuned into resonance with the excitation laser, the nuclear spin bath is clearly subject to DNP since a shift of the $\left| E_1 \right\rangle$ resonance to higher energies is observed. When the electric field is swept in the opposite direction (red trace, closed squares in Fig.~1b), first the $\left| E_1 \right\rangle$ state is tuned into resonance with the laser leading to a buildup of $B_{\mathrm{N}}$. After the $\left| E_1 \right\rangle$ state has been tuned through the laser, the energetically lower $\left| E_2 \right\rangle$ state approaches the laser energy. The measurements presented in Fig.~1b clearly show that the $\left| E_2 \right\rangle$ peak in the PC signal is now red shifted as compared to the reference measurement. This observation unequivocally shows that the nuclear field created by optical pumping of $\left| E_1  \right\rangle$ is still present.

Figure 2 presents time resolved PC measurements performed at $B_{\mathrm{ext}} = 2.5$T. These measurements allow us to investigate the buildup dynamics of $\delta_{\mathrm{N}}$. The timing and electric field sequence used for the measurements is depicted schematically in Fig.~2a. First, we employed the randomization procedure to delete any residual DNP. Following this, the electric field is set to a value $F_{\mathrm{S}}$ that defines a specific initial (red) detuning ($d_{\mathrm{i}}$) of $\left| E_1 \right\rangle$ from the laser. Finally, the laser is switched on and we record the PC signal as a function of time. The middle and right panel of Fig.~2b shows the temporal dependence of the PC for different $d_{\mathrm{i}}$. 
For all detunings $d_{\mathrm{i}} < 36\mu$eV, the PC signal first increases to a common maximum value and then decreases to a steady state value $I_{\mathrm{PC}}^{\mathrm{equ}}$ characteristic of each $d_{\mathrm{i}}$. For $d_{\mathrm{i}} > 36\mu$eV a monotonic increase in the PC signal towards the individual $I_{\mathrm{PC}}^{\mathrm{equ}}$ value is observed. Plotting $I_{\mathrm{PC}}^{\mathrm{equ}}$ versus $d_{\mathrm{i}}$ and comparing the result with the electric field sweep measurements presented in the left panel of Fig.~2b clearly shows that they directly map out the sweep-up curve (yellow/light gray symbols and blue trace, closed triangles in Fig.~2b). 
This observation directly confirms that the electric field sweeps presented in Fig.~2b are indeed performed adiabatically, such that the system remains in a steady state during measurement. Time resovled control experiments performed at different $B_{\mathrm{ext}}$ confirm that this is also the case for all electric field sweeps.
The initial increase of the PC signal arises from the buildup of an Overhauser shift that serves to reduce the red detuning of $\left| E_1 \right\rangle$ from the laser. The maximum PC signal is reached when the $\left| E_1 \right\rangle$ state is exactly in resonance with the laser. As $\delta_{\mathrm{N}}$ increases further, $\left| E_1 \right\rangle$ becomes blue detuned from the laser until the system reaches the steady state value $\delta_{\mathrm{N}}^{\mathrm{s}}$. 
These measurements allow us to directly monitor the time evolution of the Overhauser shift of $\left| E_1 \right\rangle$. To do this, we mapped the steady state PC spectrum recorded with randomization (green trace, closed circles in Fig.~2b) onto the time resolved data by shifting it such as to reproduce the instantaneous PC signal measured at a time $t$. The instantaneous Overhauser shift $\delta_{\mathrm{N}}(t)$ can then be extracted from this shift and the results of this procedure are plotted in the left panel of Fig.~2c for different values of  $d_{\mathrm{i}}$.
A characteristic behavior is observed for all curves that becomes more pronounced for larger $d_{\mathrm{i}}$. A first slow linear increase in $\delta_{\mathrm{N}}$ is followed by a faster buildup as the steep slope on the high energy flank of the resonance approaches the laser energy. This direct resonance is accompanied by a strong increase in excitation rate of the QD and, thus, a much higher nuclear spin pumping rate. 
Finally, the increase in $\delta_{\mathrm{N}}$ slows down as the system approaches the steady state. In the right panel of Fig.~2c, we plot the time required to reach saturation $t_{\mathrm{equ}}$ (Ref. 19) as a function of $d_{\mathrm{i}}$. The dependence of $t_{\mathrm{equ}}$ on $d_{\mathrm{i}}$ can be well described using a mono-exponential fit with an offset (solid black line) that reflects the asymptotic approach of the system toward the steady state. Even for initial detunings far below one linewidth ($\approx 15 \mu$eV), where the QD excitation rate is highest, we always find $t_{\mathrm{equ}} > 30$s. For $d_{\mathrm{i}} > 15\mu$eV, the initial overlap of $\left| E_1  \right\rangle$ with the laser is much smaller. The lower excitation rate results in a slower $\delta_{\mathrm{N}}$ buildup. We measured the buildup dynamics of $\delta_{\mathrm{N}}$ for $d_{\mathrm{i}}$ up to $40\mathrm{\mu eV}$, leading to $t_{\mathrm{equ}}$ in excess of 700s. For $d_{\mathrm{i}} > 40 \mu$eV, no DNP is observed over a timescale of 1000s. These measurements allow us to directly estimate the net electron-nuclear spin flip-flop rate $\Gamma_{\mathrm{ff}}$ and the efficiency of the process. At $B_{\mathrm{ext}}=2.5$T, we obtain $\delta_{\mathrm{N}}^{\mathrm{s}}=80 \mu$eV, corresponding to a nuclear spin bath polarization of $P = 32$\% \cite{Ebl06}, i.e. $\sim 10^4$ polarized nuclei. For small detunings ($ d_{\mathrm{i}} < 15 \mathrm{\mu eV}$), $t_{\mathrm{equ}}$ is typically on the order of 10s or longer indicating an average $\Gamma_{\mathrm{ff}}$ $\sim 10^3$s$^{-1}$. We note that this is an average value since the detuning and, therefore, the dynamics change continuously  throughout the measurement. In our experiment the amplitude of typical PC signals are 100-200pA, corresponding to the generation of $\sim 10^{10}$ excitons per second. These observations show that only one electron in $10^7$ actually undergoes a net spin flip-flop event with a nucleus before tunneling out of the dot.

\begin{figure}[t]
\centering
		\includegraphics[width=0.35\textwidth]{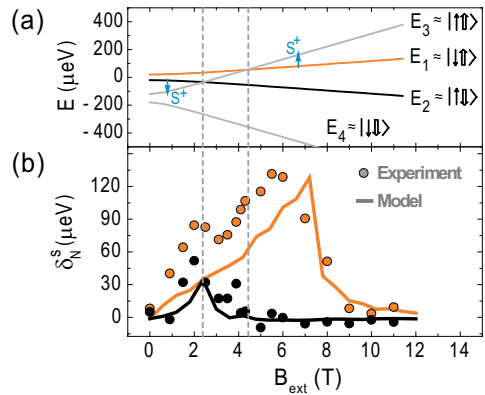}
	\label{fig:plot002}
	\vspace{-0.3cm}
	\caption{ (colour online) (a) Calculated Breit-Rabi diagram for the bright ($\left| E_{1,2} \right\rangle$) and dark ($\left| E_{3,4} \right\rangle$) exciton states. (b) Steady state Overhauser shift $\delta_{\mathrm{N}}^{\mathrm{s}}$ created by resonant excitation of either of the two X$^0$ states $\left| E_1 \right\rangle$ and $\left| E_2 \right\rangle$ as a function of $B_{\mathrm{ext}}$ obtained from experiment (circles) and calculated using the model described in the text (solid line). The dashed vertical lines indicate the crossing of $\left| E_3 \right\rangle$ and $\left| E_{1,2} \right\rangle$ in the absence of DNP, respectively.}
\end{figure}

A systematic investigation of $\delta_{\mathrm{N}}^{\mathrm{s}}$ as a function of the applied magnetic field is presented in Fig.~3b. For $B_{\mathrm{ext}}  = 0 - 6$T, $\delta_{\mathrm{N}}^{\mathrm{s}}$ created via resonant excitation of $\left| E_1 \right\rangle$ increases with increasing $B_{\mathrm{ext}}$ and then decreases monotonically for $B_{\mathrm{ext}} \geq 6$T. The maximum observed Overhauser shift of $\delta_{\mathrm{N}}^{\mathrm{s}} \approx 135 \mu$eV obtained for $B_{\mathrm{ext}} \approx 6$T corresponds to a nuclear polarization of $P = 54$\% (Ref. 7) and an Overhauser field of $B_{\mathrm{N}}=3.8$T (using $g_{\mathrm{e}} = -0.6$ calculated for our InGaAs QDs (Ref. 20) and similar to values found in the literature \cite{Kro08, Bay99}). In strong contrast, resonant excitation of $\left| E_2 \right\rangle$ does not result in any pumping of the nuclear spin bath for $B_{\mathrm{ext}} \geq 4$T. For $B_{\mathrm{ext}} \leq 4$T a small Overhauser shift is observed. However, the magnitude of $\delta_{\mathrm{N}}^{\mathrm{s}}$ is always significantly smaller than that induced by pumping of $\left| E_1 \right\rangle$ at the same $B_{\mathrm{ext}}$. Most remarkably, the direction of $B_{\mathrm{N}}$ with respect to $B_{\mathrm{ext}}$ is found to be identical for excitation of both $\left| E_1 \right\rangle$ and $\left| E_2 \right\rangle$, since in both cases $\delta_{\mathrm{N}}$ is found to result in an increase in $\Delta E$ over the value measured without DNP effects.

The pumping of the nuclear spin bath can be explained via hf-mediated electron-nuclear spin flip-flop processes that exchange the orientation of the electron and a nuclear spin. This process is described by the flip-flop part of the hf Hamiltonian $H_{\mathrm{ff}}\sim S^+\sum_j I^-_j+S^-\sum_j I^+_j$ \cite{Sch03}, where $S^\pm$ and $I^\pm_j$ are the raising and lowering operators for the electron and $j$th nuclear spin, respectively.
To understand the principle characteristics of the DNP curve in Fig.~3b we consider the exciton level structure shown in Fig.~3a. The two bright exciton states $\left| E_{1,2} \right\rangle$ are split from the two optically inactive (dark) excitons $\left| E_{3,4} \right\rangle$ by the large isotropic exchange splitting ($\sim150\mu$eV). \cite{Bay02} 
Bright and dark doublets are also split by anisotropic exchange $\delta_1$ (with $\delta_1^{\mathrm{bright}} = 40\mu$eV measured at $B_{\mathrm{ext}}=0$T) and are, therefore, a superposition of the pure spin states with a mixing ratio that decreases with increasing $B_{\mathrm{ext}}$.
The hf flip-flops couple bright and dark excitons, $ \left| \uparrow \Downarrow \right\rangle \stackrel{S^-}{\rightarrow} \left| \downarrow \Downarrow \right\rangle $ and $ \left| \downarrow \Uparrow \right\rangle \stackrel{S^+}{\rightarrow} \left| \uparrow \Uparrow \right\rangle$. Since only bright excitons are excited by the laser and tunneling quickly removes all excitons from the QD, the hf flip-flops occur only from bright to dark states. As we see from Fig.~3a the dark exciton state $\left| E_3 \right\rangle \approx \left| \uparrow \Uparrow \right\rangle$ is energetically closer to both bright states than $\left| E_4 \right\rangle \approx \left| \downarrow\Downarrow \right\rangle$. This makes hf-induced transitions to $\left| E_3 \right\rangle$ in which the electron spin flips up ($S^+$) more likely than those to $\left| E_4 \right\rangle$ in which it flips down (see Fig.~3a), since the flip rate scales with the inverse of the energy difference squared (and the matrix elements between the two states).\cite{Sak94} Consequently, the resonant optical pumping has the result that nuclear spins are predominantly flipped to align anti-parallel with $B_{\mathrm{ext}}$. This qualitatively explains why we observe \emph{unidirectional} nuclear polarization independent of the net polarization of the electrons pumped through the dot.
The peaks observed in Fig.~3b are related to the crossing of different exciton levels: as $\left| E_3 \right\rangle$ crosses either $\left| E_2 \right\rangle$ or $\left| E_1 \right\rangle$, hf flips become very likely and nuclear polarization builds up. This process is particularly efficient for the higher energy exciton $\left| E_1 \right\rangle$ (orange/dark gray curve in Fig.~3b) since the coupling matrix element $\left\langle E_3 \left| H_{\mathrm{ff}} \right| E_1 \right\rangle $ is close to a maximum as $\left| E_1 \right\rangle$ is of predominant $ \left| \downarrow \Uparrow \right\rangle$ character. Efficient coupling between $\left|E_1\right\rangle$ and $\left|E_3\right\rangle$ is preserved over a wide range of electric fields since DNP causes both an energy shift of $\left| E_1 \right \rangle$ and shifts the magnetic field for which $\left| E_1 \right \rangle$ and $\left| E_3 \right\rangle$ cross to larger values. This leads to a stronger and higher peak in the achieved nuclear polarization that is also shifted to higher $B_{\mathrm{ext}}$ values than expected from the crossing of the states without $B_{\mathrm{N}}$.
In contrast, for the lower energy exciton $\left| E_2 \right\rangle$ DNP is much less effective since this state has predominant $ \left| \uparrow \Downarrow \right\rangle$ character and $\left\langle E_3 \left| H_{\mathrm{ff}} \right| E_2 \right\rangle$ is small but non-zero due to the (weak) admixture of $\left| \downarrow \Uparrow \right\rangle$ in $\left| E_2 \right\rangle$ arising from the anisotropic exchange. This leads to the small DNP peak at low magnetic field around the crossing of $\left| E_2 \right\rangle$ and $\left| E_3 \right\rangle$ (black curve in Fig.~3b). No DNP is observed at $B_{\mathrm{ext}}\approx0$T since now both doublets are of strongly mixed spin character \cite{Bay02} and the energy difference to both dark states is almost identical for each bright state. For large $B_{\mathrm{ext}}$, hf flip-flops are ineffective since the bright and dark exciton states are strongly detuned with the result that DNP is suppressed.

All these features are reproduced by our simulations shown in Fig.~3b. Our theoretical approach makes use of the separation of time scales between the fast excitonic and slow nuclear dynamics. The exciton experiences the Overhauser field of quasi-static nuclei and quickly reaches a steady state determined by optical driving, tunneling, and the hf coupling. Dark exciton populations in this steady state arise from the transverse components of $B_{\mathrm{N}}$ (hf flip-flops) and determine (on the slow nuclear time scale) the ``instantaneous'' nuclear polarization rate. This rate is used in the simulation of the PC measurements: for given magnetic field $B_{\mathrm{ext}}$, laser frequency $\omega_{\mathrm{L}}$, and nuclear polarization $P$ we determine the rate, change $P$ accordingly and repeat until the steady state polarization is reached. Keeping $P$ fixed, the detuning is then shifted and the process is repeated. From the average exciton population (at given $B_\mathrm{ext}, \omega_{\mathrm{L}}$, and $ d_{\mathrm{i}}$) the photocurrent is deduced and from the peaks in the PC curve $\delta_{\mathrm{N}}$ is obtained as described before. In the simulations we assumed a finite laser-induced nuclear depolarization rate, as without it the nuclear steady state would be almost fully polarized throughout. A more detailed analysis of the nuclear dynamics subject to optical pumping in the presence of tunneling will be presented elsewhere.
At this point we would like to mention that we were recently made aware of a paper that describes optically mediated DNSP in a single InP/GaInP QD interrogated via photoluminescence.\cite{Che10} The model utilized to explain the results of that paper is fully consistent with the one proposed here.

This work was financially supported by the DFG via SFB 631 and NIM as well as by the EU via SOLID. \\

\clearpage

\end{document}